# Crossover in the pressure evolution of elementary distortions in $R$FeO$_3$ perovskites and its impact on their phase transition

R. Vilarinho,[1,*] P. Bouvier,[2] M. Guennou,[3] I. Peral,[3,4] M. C. Weber,[5] P. Tavares,[6] M. Mihalik jr.,[7] M. Mihalik,[7] G. Garbarino,[8] M. Mezouar,[8] J. Kreisel,[3] A. Almeida,[1] and J. Agostinho Moreira[1,*]

[1]IFIMUP, Physics and Astronomy Department, Faculty of Sciences, University of Porto, Rua do Campo Alegre 687, s/n- 4169-007 Porto, Portugal.

[2]Université Grenoble-Alpes, CNRS, Institut Néel, 38000 Grenoble, France.

[3]Materials Research and Technology Department, Luxembourg Institute of Science and Technology, 41, rue du Brill, L-4422 Belvaux, Luxembourg.

[4]Physics and Materials Science Research Unit, University of Luxembourg, 162a avenue de la Faïencerie, 1511 Luxembourg, Luxembourg

[5]Department of Materials, ETH Zurich, Vladimir-Prelog-Weg 4, 8093 Zurich, Switzerland.

[6]Centro de Química – Vila Real, Departamento de Química, Universidade de Trás-os-Montes e Alto Douro, 5000–801 Vila Real, Portugal.

[7]Institute of Experimental Physics Slovak Academy of Sciences, Watsonova 47, Košice, Slovak Republic.

[8]European Synchrotron Radiation Facility, 38043 Grenoble, France.

*Corresponding author: rvsilva@fc.up.pt

*Corresponding author: jamoreir@fc.up.pt

## 1. Abstract

This work reports on the pressure dependence of the octahedra tilts and mean Fe-O bond lengths in $R$FeO$_3$ ($R$=Nd, Sm, Eu, Gd, Tb and Dy), determined through synchrotron X-ray diffraction and Raman scattering, and their role on the pressure induced phase transition displayed by all of these compounds. For larger rare-earth cations (Nd-Sm), both anti- and in-phase octahedra tilting decrease as pressure increases, whereas the reverse behavior is observed for smaller ones (Gd-Dy). EuFeO$_3$ stands at the borderline, as the tilts are pressure independent. For the compounds where the tilts increase with pressure, the FeO$_6$ octahedra are compressed at lower rates than for those ones exhibiting opposite pressure tilt dependence. The crossover between the two opposite pressure behaviors is discussed and faced with the rules grounded on the current theoretical approaches. The similarity of the pressure-induced isostructural insulator-to-metal phase transition, observed in the whole series, point out that the tilts play a minor role in its driving mechanisms. A clear relationship between octahedra compressibility and critical pressure is ascertained.

## 2. Introduction

Hydrostatic pressure has been increasingly considered in the study of critical phenomena, since it allows to modify the interatomic distances and, consequently, the interactions to a greater extent than any other external parameter, like temperature or magnetic field. Thus, the number of published experimental[1–5] and theoretical[6–9] reports concerning the pressure evolution of elementary distortions and phase transitions in many materials has steadily increased over the last few years. In this regard, the effect of pressure on $ABO_3$ perovskites has been the focus of intense research, because of their remarkable pressure-induced phase transitions sequences, many of them undergoing many structural transformations accompanied by changes on the magnetic, transport and ferroelectric properties.[1–5]

The $ABO_3$ perovskites exhibit a rather simple crystallographic structure that can be described as a corner-sharing $BO_6$ octahedra network, with the A-cations placed between them, forming $AO_{12}$ dodecahedra.[10] The structure of $ABO_3$ perovskites can be obtained from basic distortions of the ideal $P m\bar{3}m$ cubic phase.[11,12] For perovskites with a tolerance factor less than unity, the most important distortion is characterized by short-period $BO_6$ octahedral rotations.[13] In the case of P$nma$ orthorhombic perovskites, these distortions are the in-phase and anti-phase tilts about the $[010]_{pc}$ and $[101]_{pc}$ pseudocubic directions,[13] which transform according to the $M_3^+$ and $R_4^+$ irreducible representations of the $Pm\bar{3}m$ space group, respectively.[11,12] These two distortions are the primary order parameters associated with the symmetry lowering, and present the largest amplitudes, which increase as the *A*-cation size decreases.[11] Other non-symmetry breaking distortions occur together with octahedra tilting, where the anti-parallel motion of the *A*-cations along the *z* pseudocubic direction with symmetry $X_5^+$ is the most relevant, as it bares the largest amplitude among the secondary distortions. Moreover, it couples to both tilts via a specific trilinear coupling term that provides an energy gain crucial to the stabilization of the P*nma* phase.[11,14]

In order to predict and explain the structural behavior of perovskites under high-pressure, several rules have been proposed, grounded on theoretical models and DFT calculations.[6–8] Based on the bond-valence concept, Zhao *et al.*[6] predicted that the $AO_{12}$ dodecahedra are expected to be significantly less compressible than the $BO_6$ octahedra in orthorhombic perovskites with both *A* and *B* cations having the formal charge +3 (3:3 perovskites), as it is the case of rare-earth orthoferrites ($RFeO_3$). Thus, the octahedral tilts should decrease with increasing pressure.[6] This model also evidences the correlation between the ratio of the $BO_6$ and $AO_{12}$ compressibilities and the rate of change of the octahedra tilting.[6] Moreover, the decrease of the octahedra tilting with pressure should yield a structural phase transition from the orthorhombic into a higher symmetric structure at some critical pressure.[6] This is the case of NdNiO$_3$ and LaGaO$_3$ which undergo a high-pressure structural phase transition into a rhombohedral symmetry, but with a different tilt system. As a matter of fact, these rules are not followed by many of 3:3 perovskites. For instance, new results in rare-earth chromites (*R*CrO$_3$), with small *R*-cations, show that octahedra tilting increase with pressure and some rare-earth orthoferrites undergo a pressure induced isostructural phase transition.[15,16]

In order to unravel the mechanisms underlying the distinct pressure dependences of octahedra tilting, Xiang *et al.*[7] conducted first-principles calculations on representative perovskites and proposed a set of rules governing the tilt evolution with pressure. In this framework, they simulated the pressure dependence of both in-phase and anti-phase octahedra tilts of LaFeO$_3$ and LuFeO$_3$, as border cases in the rare-earth orthoferrites series.[7] For LaFeO$_3$, they found that both octahedra tilts are suppressed as pressure increases in agreement with Zhao's prevision.[7] However, pressure suppresses the anti-phase but enhances the in-phase tilting in LuFeO$_3$.[7] This pressure behavior was interpreted by taking into account the contribution of a trilinear coupling between these two rotations and the anti-polar mode involving the *A*-cation.[7] Thus, a new rule emerges for the orthorhombic P*nma* perovskites, which simultaneously exhibit the in-phase and anti-phase octahedra tilts, stating that they are not inevitably both suppressed or enhanced by pressure.[7] Unfortunately, the authors did not present results concerning intermediate rare-earth orthoferrites in order to check where the crossover between the two aforementioned distinct pressure behaviors occurs.

Despite the intensive research already done,[16–21] a systematic study of the pressure evolution of the elementary structural distortions in orthorhombic perovskites is still missing, as well as, the careful search for crossover events, including the experimental determination of the borderline compound. In this regard, the study of the *R*FeO$_3$ series under high pressure is particularly interesting, as it is expected to show different pressure dependences of the elementary distortions across the series.[7,8] Moreover, this system bears a distinct advantage, as it does not show distortions other than those with direct origin in octahedra tilting, like Jahn-Teller distortion, which can influence the mean B-O bond length.[2]

In this work, we present an experimental study of the structure of the *R*FeO$_3$ (*R* = Nd, Sm, Eu, Gd, Tb and Dy) as a function of pressure, by means of synchrotron X-ray diffraction and Raman scattering. We first analyze the pressure evolution of the octahedra tilt angles across the series in order to reach an overall picture of its dependence on the rare-earth cation size. Then, the pressure variation of the mean Fe-O bond length obtained by Raman scattering is explored in order to correlate the pressure tilt behavior and the mechanisms to accommodate pressure. We also discuss the experimental results within the scope of the theoretical models' predictions. Finally, the dependence of the critical pressure on the rare-earth cation size is examined in terms of the pressure behavior of the elementary distortions.

## 2. Experiment aspects

SmFeO$_3$ single crystals were grown in an optical-floating-zone furnace,[22] and NdFeO3 and TbFeO$_3$ powder was prepared using single crystals growth by floating zone method in a FZ-T-4000 (Crystal Systems Corporation) mirror furnace. EuFeO$_3$ powder was obtained by conventional solid-state reactions, while GdFeO$_3$ and DyFeO$_3$ powder was prepared using the urea sol-gel combustion method.[23] The quality of the samples was previously characterized by means of X-ray diffraction, Fourier transform infrared spectroscopy, X-ray photoemission spectroscopy and scanning electron microscopy. The ceramics were manually grinded to acquire a homogeneous powder and loaded in a diamond anvil cell (DAC) with diamond culets of 300 μm diameter, and with helium as a pressure-transmitting medium. The pressure was monitored through the standard fluorescence method of a ruby loaded next to the sample. High-pressure synchrotron X-ray diffraction (XRD) experiments on SmFeO$_3$ and TbFeO$_3$ were performed at the European Synchrotron Radiation Facility (ESRF, Proposal Ref. HC-2153) on the ID27 high-pressure beamline (λ=0.3738 Å). The diffraction data were analyzed by Le Bail and Amplimodes refinements using FullProf and Jana2006 softwares. The Raman spectra were recorded on a Horiba LabRam at MINATEC (Grenoble, France) using a He-Ne laser at 633 nm for TbFeO$_3$ and on a Horiba T64000 spectrometer at Institut Néel (Grenoble, France) using an Ar+ laser at 514.5nm for *R*FeO$_3$ (*R*=Nd, Sm, Eu, Gd and Dy). In both cases, the laser power was kept below 8 mW on the DAC to avoid sample heating. Raman spectra were fitted by a sum of independent damped oscillators using IgorPro® software.

## 3. Results

### a. High pressure XRD: SmFeO$_3$ and TbFeO$_3$

Figure 1 shows the pressure dependence of the pseudocubic lattice parameters $a_{pc} = a/\sqrt{2}$, $b_{pc} = b/2$, and $c_{pc} = c/\sqrt{2}$ (values can be found in Table I of *Supplemental Material*), and the pseudocubic volume $V_{pc} = V/4$ ($Z = 1$), with $V$ the volume of the primitive cell of the P*nma* structure of SmFeO$_3$ and TbFeO$_3$. See Figure S1 of *Supplemental Material*, where representative XRD patterns of both compounds, recorded at different pressures, are shown.

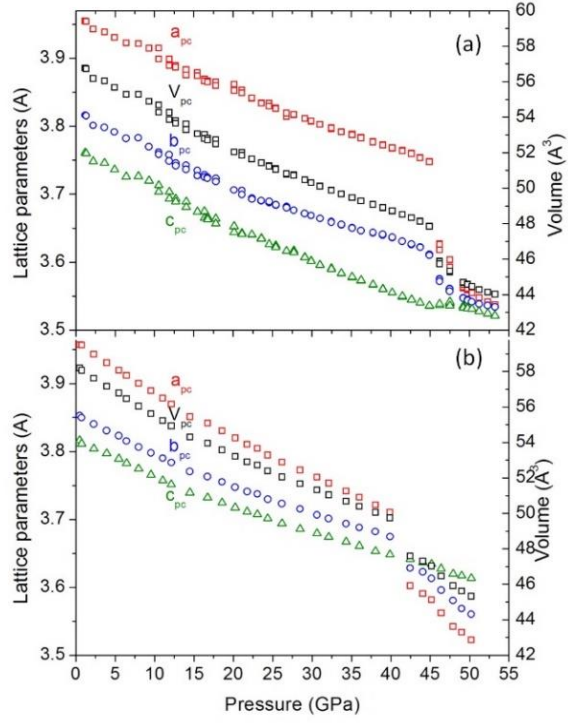

**Figure 1.** Pseudocubic lattice parameters and volume as a function of pressure for (a) TbFeO$_3$ and (b) SmFeO$_3$.

As the pressure increases, the pseudocubic lattice parameters monotonously decrease and suddenly down shift at around 41 GPa for SmFeO$_3$, and at 46 GPa for TbFeO$_3$; as a consequence, the pseudocubic volume is reduced 4.3% and 6.6%, respectively. The later result gives clear evidence for a pressure-induced structural phase transition, which we will address in detail in the last section of this work. On further pressure increase, a smooth pressure evolution of the pseudocubic lattice parameters, and volume is again observed. The symmetry of both low- and high-pressure phases of SmFeO$_3$ and TbFeO$_3$ is found to be P*nma*. This symmetry was reported for the high pressure phase of other rare-earth orthoferrites.[16,19]

In the low pressure phase, the pressure dependence of the pseudocubic cell volume of SmFeO$_3$ and TbFeO$_3$ can be adequately described by the third-order Birch-Murnaghan isothermal equation of state:[24,25]

$$P = 3B_o f_E [1 + 2f_E]^{5/2} \left\{1 + \frac{3}{4}(B'_o - 4)f_E\right\} \qquad (1)$$

where $B_0$ the bulk modulus and $B'_0$ its pressure derivative, all taken at room pressure, and $f_E$ is given by:

$$f_E = \frac{1}{2}\left\{\left(\frac{V_{pc}(0)}{V_{pc}}\right)^{2/3} - 1\right\} \qquad (2)$$

where $V_{pc}$ is the pseudo-cubic volume at the pressure $P$, and $V_{pc}(0)$ its room pressure value. Figure S2 of the *Supplemental Material* shows the best fit of Equation (1) to the experimental data. Table 1 presents the values of $B_0$ and $B'_0$, obtained from the fit procedure. For both compounds, the values of the bulk modulus are not much different, taking values around 182 GPa, while the $B'_0$ is about 4, typical of nearly isotropic compressions.

**Table 1.** Pseudocubic volume at room pressure $V_{pc}(0)$, bulk modulus $B_o$, and its first pressure derivative $B'_0$, obtained from the best fit of the third-order Birch-Murnaghan equation of state (Eq. 1) to the pseudocubic volume of TbFeO$_3$ and SmFeO$_3$.

|  | $V_{pc}(0)$ (Å³) | $B_o$ (GPa) | $B'_0$ |
|---|---|---|---|
| SmFeO$_3$ | 58.3±0.1 | 183±3 | 4.0±0.2 |
| TbFeO$_3$ | 57.2±0.1 | 181±9 | 4.0±0.6 |

### b. Raman scattering

According to group theory, we expect 24 Raman-active vibration modes for the orthorhombic P*nma* space group (see Ref. 3 for the symmetry of the Raman-active modes). The Raman spectra of *R*FeO$_3$, with *R* = Nd, Sm, Eu, Gd, Tb and Dy, recorded at different applied pressures are shown in Figure S3 of *Supplemental Material*. They exhibit simultaneously all Raman-active modes due to the unpolarized recording condition. A detailed mode assignment of the Raman bands and the corresponding atomic motions for rare-earth orthoferrites is presented elsewhere.[3]

The Raman spectra for all the studied materials exhibit similar trends with increasing pressure: Raman bands shift towards higher wavenumbers, due to an overall pressure-induced bond shortening and volume reduction, and become broader; their intensity reduces and disappear above a certain pressure that depends on the compound. The later result corroborates the existence of a structural phase transition at high pressures for all the studied compounds. The nature and structure of the high-pressure phase will be discussed in the last section.

In the following, we focus our attention on the Raman-active modes assigned to the FeO$_6$ octahedra rotations, mirroring the anti-phase and in-phase octahedra tilts. Figure 2 presents the pressure dependence of the wavenumber of the aforementioned Raman modes of the studied compounds. The pressure dependence of the wavenumber of the Raman modes does not exhibit anomalous behavior up to the critical pressure, corroborating that the P*nma* structure is preserved in the low-pressure range. A linear pressure dependence of the wavenumber of the Raman bands is observed below to 20 GPa. Concerning NdFeO$_3$, our results agree with those already reported below 11 GPa.[21] From the best fit of a linear function to the experimental data below to 20 GPa, shown in Figure 2, we have determined the corresponding slopes and wavenumbers at room conditions which are presented in Table 2.

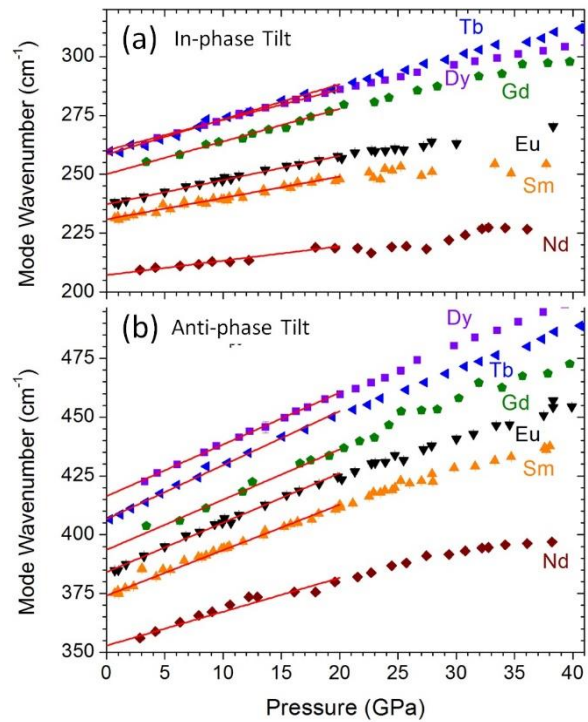

**Figure 2.** Pressure evolution of the Raman modes wavenumber associated with (a) in-phase and (b) anti-phase octahedra rotations for the different studied $R$FeO$_3$. The lines were obtained from best linear fits to the data recorded below 20 GPa.

The slopes of the linear pressure dependence of the wavenumber of both in-phase and anti-phase tilt modes tend to increase as the rare-earth cation size decreases. We can see this result in another way: for larger tilt angles measured at ambient pressure, greater slopes are observed, as it is predicted by theoretical models.[7]

**Table 2**. Wavenumbers at room pressure, and slopes of the linear pressure dependence of the in-phase and anti-phase tilt modes, involving the data recorded below to 20 GPa.

| Compound | In-phase (cm$^{-1}$/GPa) | | Anti-phase (cm$^{-1}$/GPa) | |
|---|---|---|---|---|
| | Wavenumber (cm$^{-1}$) | Slope (cm$^{-1}$/Gpa) | Wavenumber (cm$^{-1}$) | Slope (cm$^{-1}$/Gpa) |
| NdFeO$_3$ | 209.4±0.8 | 0.43±0.08 | 356.0±0.2 | 1.84±0.07 |
| SmFeO$_3$ | 231.2±0.2 | 0.91±0.04 | 375.2±0.2 | 2.10±0.05 |
| EuFeO$_3$ | 238.0±0.1 | 1.10±0.05 | 384.8±0.1 | 2.15±0.05 |
| GdFeO$_3$ | 255.2±0.3 | 1.34±0.03 | 403.7±0.3 | 2.04±0.04 |
| TbFeO$_3$ | 259.9±0.4 | 1.53±0.05 | 406.5±0.4 | 2.43±0.05 |
| DyFeO$_3$ | 262.5±0.1 | 1.43±0.04 | 422.5±0.1 | 2.28±0.05 |

## 4. Discussion

### a. Compilation and analysis of XRD data

The quality of the XRD patterns obtained at high pressures hinders the full Rietveld refinement of the atomic positions and, so, the calculation of the tilt angles. This is however possible with the Amplimodes analysis.[11,26] During the structure refinement, instead of allowing the atomic positions to vary in the three dimensional space without restriction to find the global minimum, Amplimodes are used to improve the structure refinement by describing the displacement of the atoms, relatively to their positions in the high symmetry $P m\bar{3} m$ structure, as the superposition of symmetry-adapted distortions.[26] The refinements show that the internal octahedral distortions have small amplitudes (less than 0.05 Å), whereas the octahedral tilts distortions and the rare-earth shifts are apparently larger (between 1 and 2 Å value of global amplitude). However, the refined mode amplitudes when plotted versus pressure show dispersion, especially in the modes mainly involving oxygen motions (that is, the tilting modes – see Figure S4 of *Supplemental Material*). This is not surprising since high-pressure powder diffraction presents peak overlapping and the contribution of the oxygen atoms to the diffraction peaks intensity is small compared to the rest of the atoms (that is Sm or Tb, and Fe). The Amplimodes refinement results, especially regarding the octahedral tilts, should be contrasted with other observations.

Thus, we estimated the tilt angles from the lattice parameters, using the equations of Megaw *et al*.[27,28] This formula assumes that the change of the unit cell volume is originated by the octahedral tilting, with no significant additional octahedra distortion.[27,28] The formula is most suitable for larger tilts than for smaller tilts, as for the latter other contributions to the change of the unit cell volume may be comparable to the contribution from the tilts.[27,28] Therefore, for our compounds, the estimation using Megaw's formula is better for the anti-phase tilt than for the in-phase tilt, and for smaller tolerance factors than for larger ones. The reliability of this approach was assessed by comparing the values of the tilt angles for these compounds, obtained at room conditions using the Megaw's formula, with their reference ones, obtained from the refinement of the atomic positions, as shown in Figure 3. Furthermore, their pressure behaviors follow similar trends as the ones found by the Amplimodes analysis, though with much less dispersion (see Figure S4 in *Supplemental Material*). Moreover, as we shall discuss in the following, this approach is also consistent with the pressure dependence of the spontaneous $e_4$ strain, calculated using the refined lattice parameters (see Figure S5 in *Supplemental Material*), which is inherently connected to the two tilts.[13] The good agreement between the literature and our experimental observations ensures the validity of this method for other compounds where similar conditions are expected, such as $LuFeO_3$, $EuFeO_3$ and $NdFeO_3$. For the these compounds, we have calculated the tilt angles from the lattice parameters measured at different pressures already published,[16,18] as these values are relevant for the discussion of how their pressure behavior depends on the rare-earth ionic radius (see following section).

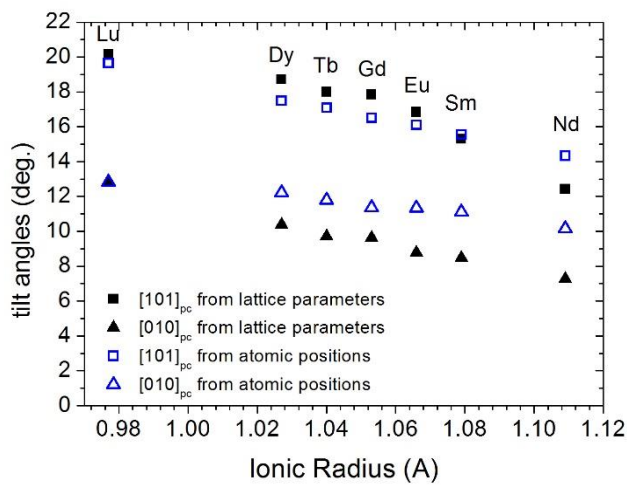

**Figure 3.** Comparison of tilt angles estimated from the lattice parameters, following the equations in Ref. [27] (closed symbols), with reference values obtained from the atomic positions (open symbols).

The pressure dependence of the anti-phase tilt angle for $R$ = Nd, Sm, Eu, Tb and Lu, and of the in-phase tilt angle for $R$ = Sm, Eu and Tb are shown in Figures 4(a) and (b), respectively. The data regarding the in-phase tilt angle for $LuFeO_3$ and $NdFeO_3$ are not presented because they are too scattered. For the same compound, both anti-phase and in-phase tilt angles present similar pressure behaviors, but the pressure trend depends on the $R$-cation. For the compounds with larger rare-earth cations ($SmFeO_3$ and $NdFeO_3$), the applied pressure causes a decrease of both anti-phase and in-phase tilt angles, and thus a decrease of the values of atomic displacements associated with the symmetry-adapted $R_4^+$ and $M_3^+$ distortion modes, while the opposite behavior is observed for the compounds with smaller rare-earth cations ($LuFeO_3$ and $TbFeO_3$). A similar trend is ascertained for the pressure dependence of the spontaneous $e_4$ strain in the same pressure range (see Figure S5 of *Supplemental Material*). The $EuFeO_3$ case sits in the transition between these two opposite pressure behaviors, where both tilt angles, and consequently, the spontaneous $e_4$ strain are almost pressure independent. Our experimental results point out that pressure weakens both octahedra tilts for compounds with larger rare-earth cations towards a less distorted structure, while it enhances them for compounds with smaller rare-earth cations. Moreover, a remarkable continuous evolution between the two opposite pressure behaviors is experimentally observed.

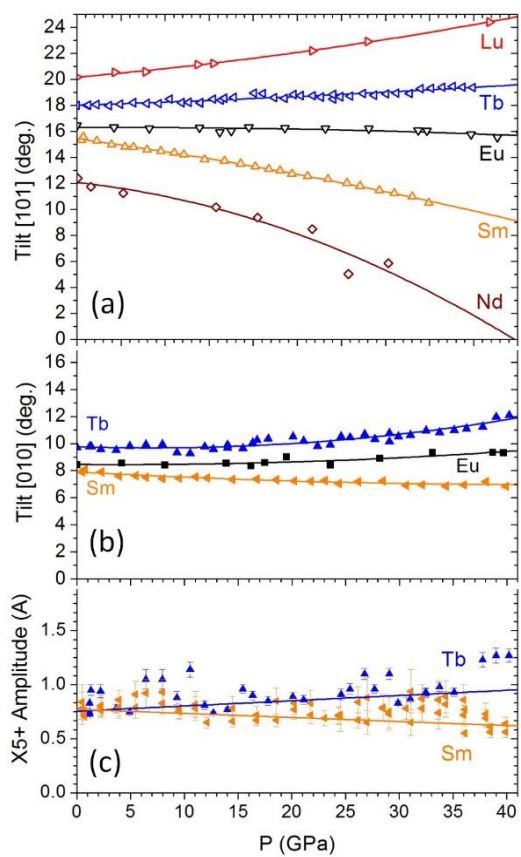

**Figure 4.** Pressure dependence of the (a) anti-phase, (b) in-phase octahedra tilts, calculated from the lattice parameters using the Megaw's formula,[27] for different ReFeO$_3$, and (c) X$_5^+$ distortion calculated from Amplimodes refinement tool for SmFeO$_3$ and TbFeO$_3$. The solid lines are guides for the eyes. Data for LuFeO$_3$ and EuFeO$_3$ from [16] and NdFeO$_3$ from [18].

As stated by Xiang *et al.*,[7] the stabilization of the P*nma* phase of perovskites is predicted to be due to an energy gain coming from a trilinear coupling of both R$_4^+$ and M$_3^+$ distortions with the X$_5^+$ distortion, associated with the displacement of the rare-earth cation.[8,14] In this theoretical framework, the magnitude of the X$_5^+$ distortion mode is predicted to be proportional to the product of both tilt distortions.[8] If this is the case, then this distortion should follow the same trend as the tilts to which it is coupled with. In fact, the pressure evolution of the X$_5^+$ distortion amplitude, which we have experimentally obtained by the Amplimodes analysis, is consistent with this model, since it mimics the behavior of the tilts for the respective rare-earth, as seen in Figure 4(c).

### b. Relation between distortion amplitudes and Raman wavenumbers

Raman scattering was also used to probe the pressure dependence of both tilt angles and octahedra distortions. For the same *RB*O$_3$ system, with fixed *B* atom, where quite similar mean B-O bond lengths are experimentally evidenced, a clear dependence of the tilt mode wavenumbers on the corresponding tilt angles was proposed by Iliev *et al.*[29] for orthomanganites, and more recently for rare-earth orthoferrites and orthochromites by Weber *et al.*[3,30] However, the tilt mode wavenumbers are also dependent on the mean B-O bond length, as it was established by Todorov *et al.*,[31] and more recently by Vilarinho *et al.*[32] So, the tilt mode wavenumbers are functions of both the tilt angle and the B-O bond length. In a general approach, the wavenumber of the tilt mode for $R$FeO$_3$ can be written as follows:[31]

$$\omega = (\alpha_1 - \alpha_2 \langle Fe - O \rangle)\varphi = (m_1 + m_2)\varphi \qquad (3)$$

where $\varphi$ is the tilt angle value, $\langle Fe - O \rangle$ is the mean Fe-O bond length, $\alpha_1 = 109.1$ cm$^{-1}$/deg, and $\alpha_2 = 42.3$ cm$^{-1}$/(Å.deg) (values taken from Ref. [31]). So, for each tilt mode, we must take into account that the slope of the pressure dependence of the corresponding wavenumber has two contributions: $m_{total} = m_1 + m_2$, where $m_1$ stands for the contribution coming from the actual tilt angle change with pressure ($\alpha_1$), and $m_2$ is the contribution coming from the isotropic reduction of the FeO$_6$ octahedra volume; i.e., of the average Fe-O distance ($\alpha_2$<Fe-O>).[31,32] In the following, we present an estimative of the $m_2$ value, and how $m_1$ changes with the rare-earth ionic radius, thus gaining insight on how the mean Fe-O bond length changes with pressure for each rare-earth orthoferrite.

The rare-earth cation size dependence of the slopes of the linear pressure relation, in the 0 – 20 GPa range, of the Raman rotation modes, and of the anti-phase and in-phase tilt angles, are depicted in Figures 5(a) and 5(b), respectively.

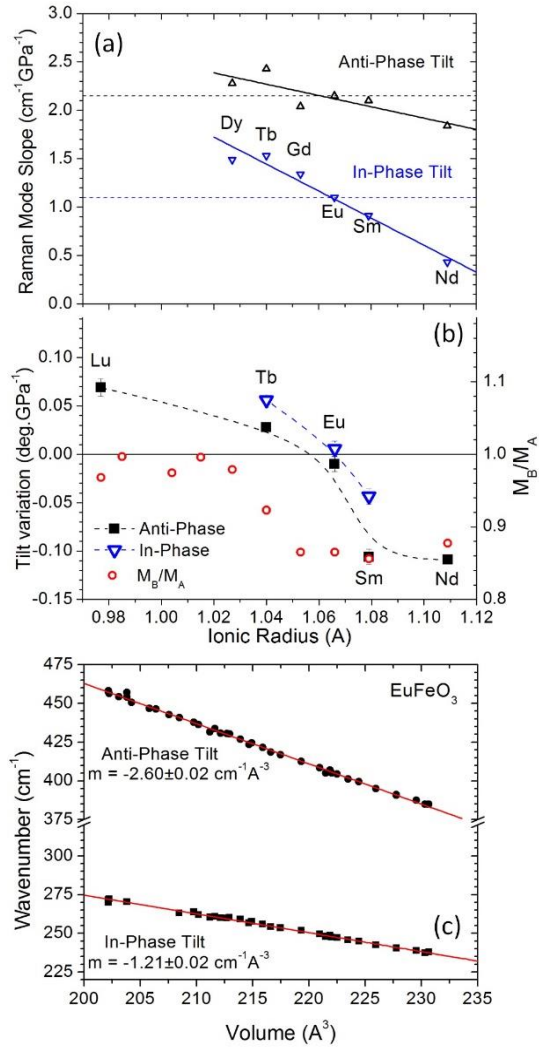

**Figure 5.** Rare-earth cation size dependence of the slopes of the linear pressure relation, in the 0 – 20 GPa range, of the (a) the Raman rotation modes and (b) of the anti-phase and in-phase tilt angles. (c) Volume dependence of the wavenumber of the Raman octahedra rotational modes for $EuFeO_3$. Dashed lines in (a) mark the reference value of $EuFeO_3$ for the contribution of the $FeO_6$ octahedra reduction. For comparison, we also present in (b) the ratio between the compressibilities of the $AO_{12}$ dodecahedra $\beta_A$ and the $FeO_6$ octahedra $\beta_B$: $\beta_A/\beta_B = M_B/M_A$, from Zhao's work.[6]

The slopes of the linear pressure dependence of the Raman rotation modes, presented in Figure 5(a), are always positive. The case of $EuFeO_3$ is straightforward because the tilt angles are pressure independent, and thus, $m_1 = 0$ (see Eq. 3). So, the linear pressure dependence of both Raman rotation modes in $EuFeO_3$, observed in the 0 – 20 GPa range, comes from the reduction of the <Fe-O> value, and $m_2 = m_{total}$, which is represented by the horizontal dashed lines in Figure 5(a). In this case, the wavenumber of the rotational modes mirrors the reduction of the unit cell volume, due to the decrease of the mean B-O bond lengths with pressure, and a linear relation between the mode wavenumber and volume is experimentally evidenced, as shown in Figure 5(c) (similar plots for TbFeO$_3$ and SmFeO$_3$ are presented in Figure S6 of *Supplemental Material* for which the linear relation does not hold, as the tilts are not constant for these compounds).

In order to unravel the effect of the tilt angles on the pressure dependence of the Raman tilt mode wavenumber for the other compounds, the $m_2 = m_{total}$ value calculated for EuFeO$_3$ might not the best reference, since the reduction of the FeO$_6$ octahedra volume may have different values for each compound, as it will be shown. According to Eq. 3, the pressure derivative of the wavenumber of the Raman tilt modes is:

$$\frac{d\omega}{dP} = [\alpha_1 - \alpha_2 \langle Fe - O \rangle(P)] \frac{d\varphi}{dP} - \alpha_2 \varphi(P) \left(\frac{d\langle Fe-O\rangle}{dP}\right). \tag{4}$$

In the following, we shall consider the values concerning $\frac{d\omega}{dP}$ and $\frac{d\varphi}{dP}$ as the slopes obtained in the 0 – 20 GPa range, shown in Figures 5(a) and 5(b), respectively. For EuFeO$_3$, where $\frac{d\varphi}{dP} = 0$, we obtained $\frac{d\langle Fe-O\rangle}{dP} = -0.0031 \pm 0.0001$ Å/GPa, using either anti- or in-phase tilts. For the remaining compounds, we focus our calculation to the linear range in the vicinity of P = 0 GPa, thus taking the values of $\langle Fe - O \rangle$ and $\varphi$ at that point. The obtained results for the low-pressure range are shown in Table 3. The estimated $\frac{d\langle Fe-O\rangle}{dP}$ values varied only around 0.001 Å/GPa when using the two different rotation modes. It is worth to note that, Weber *et al.*[3] have found different proportionality constants ($m_{total}$) between ω and φ for the anti- and in-phase tilts for the RFeO$_3$ system. The presented value agrees with that obtained by Todorov *et al.*[31] for the anti-phase tilt, but not for the in-phase tilt. Therefore, the most suitable value for the actual $\frac{d\langle Fe-O\rangle}{dP}$ should be the one obtained with the anti-phase tilt.

Table 3. d<Fe-O>/dP calculated from Eq. 4, for the low-pressure range, using either the anti- and the in-phase tilt angles, for the different rare-earth orthoferrites.

| Tilt angle | d<Fe-O>/dP (Å/GPa) | |
|---|---|---|
| | Anti-Phase | In-Phase |
| TbFeO$_3$ | -0.0023 ± 0.0003 | -0.0009 ± 0.0004 |
| EuFeO$_3$ | -0.0031 ± 0.0001 | -0.0031 ± 0.0001 |
| SmFeO$_3$ | -0.0071 ± 0.0003 | -0.0058 ± 0.0006 |
| NdFeO$_3$ | -0.0082 ± 0.0003 | N/A |

These results evidence that, at low pressures, as the rare-earth ionic radius increases, the pressure rate at which the FeO$_6$ octahedra reduce also increases. For TbFeO$_3$, the average reduction of the Fe-O mean bond length is of -0.0016 Å/GPa, half the value of EuFeO$_3$, and around 5 times smaller than the largest one of -0.0082 Å/GPa for $NdFeO_3$. This is not unexpected, since the tilts provide a mechanism of pressure accommodation. Thus, for the smaller rare-earth cations, where the tilts increase with pressure, the $FeO_6$ octahedra are compressed at lower rates. Conversely, for the larger rare-earths, where the tilts decrease with pressure, the $FeO_6$ octahedra are compressed at larger rates.

### c. Crossover for the behavior of tilts in the rare-earth series: experiment and theory

We now focus on the evolution of the tilt angles in the vicinity of zero (atmospheric) pressure across the series. From Figure 5(b), we can observe that the slope of the linear pressure dependence of both anti-phase and in-phase octahedra tilt angles is positive for compounds with smaller rare-earth cations ($R$ = Lu and Tb), while they are negative for those with larger rare-earth cations ($R$ = Sm and Nd). $EuFeO_3$ stands at the borderline, as for this compound both slopes are negligibly small. The modulus of the slope of the linear pressure dependence of the in-phase tilt angle is smaller than the anti-phase tilt angle for $SmFeO_3$, while for $TbFeO_3$ it is the opposite. These results show that the way the pressure is accommodated is different for compounds with different rare-earth cationic sizes.

We now compare this compilation of experimental results with predictions made by theoretical approaches for the pressure evolution of tilt angles. A first approach is based on the compressibility of the polyhedra forming the perovskite structure. In this approach, the ratio of compressibilities $ß_A/ß_B$ between the $AO_{12}$ and $BO_6$ polyhedra is used as a predictor of the behavior of tilt angles under pressure, and is calculated by a valence bond sum model: tilt angles are expected to increase when this ratio is larger than unity, and decrease otherwise. The case of the rare-earth orthoferrites has been treated in Ref. 6, and the corresponding data are reported in Figure 5(b) for comparison. The ratio is always positive, but decreases as the ionic radius increases and reaches values very close to unity for the smallest cations.[6] According to this criterion, all compounds in the series should see their tilt angles reduced under pressure, with the exception of $TmFeO_3$.[6] This is not what is found experimentally, but the overall evolution bears a striking resemblance with the experimental one, up to an overall shift.

More recently, in a DFT-based approach, a set of rules for the evolution of tilts under pressure was proposed.[7] In this paper, the compounds with the smallest *A*-cations (Lu, Tm), are predicted to exhibit an unusual behavior whereby the two tilt angles behave in the opposite way with pressure: the antiphase tilts are reduced whereas the in-phase tilts are enhanced.[7] For larger cations, both tilts behave the same way and are reduced as pressure increases. We find here no evidence for this behavior, but instead, both tilt angles behave the same in all investigate compounds. Also, Ref. 7 does not predict any case where both tilts increase under pressure.

The discrepancies between theoretical and experimental results seemingly point to a specific difficulty in predicting the tilt behavior for the smallest cations in the series. While we cannot be conclusive at this point about the precise reasons for this discrepancy, we observe that, in both approaches, the predicted tilt changes under pressure reach extremely small values, typically below 0.01°/GPa for $LuFeO_3$ in Ref. 7. For $EuFeO_3$, which in fact has absolute pressure derivatives below 0.01°/GPa, the slopes for in-phase and anti-phase tilts are of opposite sign (0.006 and -0.01°/GPa, respectively), but with values negligibly small that fall within the experimental uncertainty. In that context, it is conceivable that some additional parameters, legitimately neglected for large cations, become relevant and suffice to change the trend from slightly positive to negative or vice versa. In particular, $RFeO_3$ with very small cations (Er-Lu) are known to exhibit some distortions of the $AO_{12}$ polyhedra, so that the classical separation of the 12 A-O bonds into 4 longer bonds and 8 shorter bonds becomes less satisfactory.[33] We hypothesize that this has an influence of the bond valence sum and compressibilities calculated in Ref. 7. In the DFT approach, where calculations are performed at 0 K, and the behavior of tilts is the result of a delicate balance between the pressure evolution of the Landau parameters, one might question the role of temperature effects. Differences in tilt angles of the order of 0.1° between 0 and 300 K – which is reasonable following Ref. [34] – would be equivalent to several tens of GPa and cause significant shifts in the predicted behavior. Altogether, further work will be needed to clarify the picture for small cations, which calls for detailed experimental studies and reexamination of theoretical models.

### d. High-pressure phase transition

Despite the crossover between enhancement and suppression of the tilt angles on the low-pressure regime, for all the studied compounds a similar isostructural pressure-induced structural phase transition occurs. This implies that the tilts play no major role on triggering this phase transition, nor on the symmetry of the high-pressure phase. This phase transition is clearly evidenced by the changes on the XRD patterns, mirroring the sudden changes of the pseudocubic lattice parameters. The structural phase transition reveals itself by the disappearance of the Raman signal above a certain critical pressure, hereafter designated by $P_{IM}$. The crystallographic structure of the high-pressure phase of $SmFeO_3$ and $TbFeO_3$ is P*nma*, as it has been reported for other rare-earth orthoferrites.[16] This structure allows for Raman signal. Therefore, the disappearance of the Raman spectra above the critical pressure suggests the metallic character of the high-pressure structural phase.

The pressure hysteresis evidenced by the XRD patterns of SmFeO$_3$ and TbFeO$_3$, and Raman spectra, for all studied compounds, recorded on increasing and decreasing pressure runs (3 to 6 GPa depending on the rare-earth cation), evidences for the first-order character of the high-pressure structural phase. Taking into account both the XRD and Raman data, we can estimate the critical pressure P$_{IM}$ for the studied compounds, which is depicted in Figure 6. As it can be observed, the critical pressure linearly increases as the rare-earth cation size decreases.

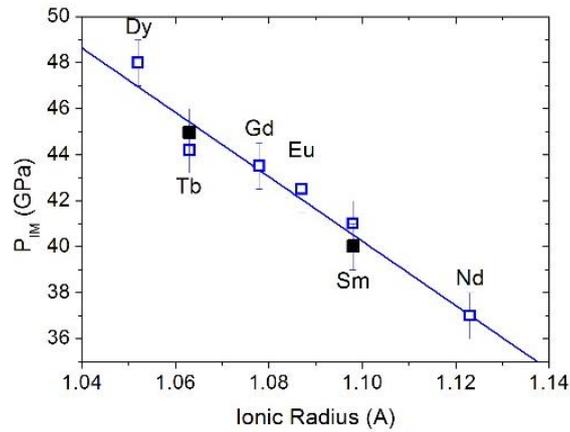

**Figure 6.** Critical pressure P$_{IM}$, obtained from XRD (closed symbols) and Raman scattering (open symbols) data, as a function of rare-earth ionic radius.

High-pressure Mössbauer experiments give evidence for a transition from high-spin to low-spin configurations occurring at P$_{IM}$.[19] As the tilts have little impact on this transition, it is sensible to assume that is depends mostly on the chemistry found inside the FeO$_6$ octahedra. We suggest that, as pressure increases and the FeO$_6$ octahedra become smaller, a critical volume of the octahedron is reached where the electronic repulsion between the oxygens p-electrons and the iron e$_g$-electrons is such, that it triggers this electronic reconfiguration, as it is energetically favorable for the e$_g$ -electrons to pair with the t$_{2g}$-electrons, avoiding to come closer to the oxygen p-electrons. The critical volume can be estimated for EuFeO$_3$. As the tilt angles do not change with pressure, there is an isotropic volume reduction of the unit cell, already evidenced in Figure 5(c), and thus the ratio of the FeO$_6$ octahedron with the pseudocubic unit cell volume (V$_{FeO6}$/V$_{pc}$) can be assumed pressure-independent. Considering a regular octahedron at room pressure, the ratio V$_{FeO6}$/V$_{pc}$ = 0.189 is obtained. Using the value of V$_{pc}$ before the critical pressure, the critical volume for EuFeO$_3$ is estimated to be V$_{FeO6}$ = 9.5±0.2 Å$^3$.

It is expected that for larger *R*-cations, where from Table 3 we know the FeO$_6$ octahedra reduce their volume at a larger rate, the aforementioned phenomenon is promoted at a lower critical pressure. Conversely, for the smaller *R*-cations, as the FeO$_6$ octahedra reduce their volume at a lower rate, they present higher values of critical pressure. This prediction explains the behavior of P$_{IM}$ with the rare-earth size, shown in Figure 6. Moreover, one can infer that a similar role is played by the octahedra tilting in the similar pressure-driven phase transition observed in *R*MnO$_3$.[2] The critical pressures for *R*MnO$_3$ have similar dependence on the rare-earth ionic size, being always slightly higher by a constant value of 2 GPa. This difference can be assigned to the Jahn-Teller distortion present in *R*MnO$_3$, as it provides an additional pressure accommodation mechanism than their respective *R*FeO$_3$ compounds. It is worth to note that the tilt angles are almost the same, thus allowing for this comparison.[32] This fact also supports the importance of the chemistry inside the octahedra in determining the symmetry of the high-pressure phase. This is because, unlike the isostructural phase transition found for the *R*FeO$_3$, the *R*MnO$_3$ present many different symmetries of the high-pressure phase.[2]

## 5. Conclusions

In this work, we present an experimental study of the pressure dependence of the main structural distortions and lattice dynamics in the *R*FeO$_3$, by XRD and Raman spectroscopy. Firstly, we have ascertained the crossover of the pressure dependence of both anti- and in-phase octahedra tilting in *R*FeO$_3$ series. EuFeO$_3$ stands distinctly at the borderline, where both tilts are pressure independent. For larger rare-earth cations, we have found that the octahedra tilting decreases as pressure increases, whereas the reverse behavior is observed for smaller ones. Furthermore, the pressure dependence of the mean Fe-O bond length, estimated from Raman data, enabled us to determine the way the pressure is accommodated in the *R*FeO$_3$ series. We observed, that for the compounds where the tilts increase with pressure, the FeO$_6$ octahedra are compressed at lower rates than for those ones showing opposite pressure tilt dependence. Finally, we determined that all the compounds of the *R*FeO$_3$ series undergo similar a pressure-induced isostructural insulator-to-metal phase transition. Thus, we have to conclude that the different pressure evolutions of the octahedra tilts play a minor role in its driving mechanisms. Moreover, we have interpreted the rare-earth ionic size dependence of the transition pressure in terms of reaching a critical volume size of the FeO$_6$ octahedra, estimated to be $V_{FeO6}$ = 9.5±0.2 Å$^3$ in EuFeO$_3$. For the smaller *R*-cations, as the FeO$_6$ octahedra reduce their volume at a lower rate, a shift of the transition to higher pressures occurs, contrarily to the case of larger *R*-cations, wherein the volume increases at a higher rate.

**Acknowledgements**

The Authors thank J. Jacobs for He gas loading at ESRF. The Authors also thank G. KH. Rozenberg and M. P. Pasternak for giving access to the details of their published XRD data concerning $EuFeO_3$ and $LuFeO_3$. The authors would like to acknowledge the support of the projects Norte-070124-FEDER-000070, PTDC/Fis–NAN/0533/2012, VEGA 2/0137/19, and R. Vilarinho to the grant PD/BD/114456/2016 by FCT.

# Crossover in the pressure evolution of elementary distortions in RFeO$_3$ perovskites and its impact on their phase transition

R. Vilarinho,[1,*] P. Bouvier,[2] M. Guennou,[3] I. Peral,[3,4] M. C. Weber,[5] P. Tavares,[6] M. Mihalik jr.,[7] M. Mihalik,[7] G. Garbarino,[8] M. Mezouar,[8] J. Kreisel,[3] A. Almeida,[1] and J. Agostinho Moreira[1,*]

[1]IFIMUP, Physics and Astronomy Department, Faculty of Sciences, University of Porto, Rua do Campo Alegre 687, s/n- 4169-007 Porto, Portugal.

[2]Université Grenoble-Alpes, CNRS, Institut Néel, 38000 Grenoble, France.

[3]Materials Research and Technology Department, Luxembourg Institute of Science and Technology, 41, rue du Brill, L-4422 Belvaux, Luxembourg.

[4]Physics and Materials Science Research Unit, University of Luxembourg, 162a avenue de la Faïencerie, 1511 Luxembourg, Luxembourg

[5]Department of Materials, ETH Zurich, Vladimir-Prelog-Weg 4, 8093 Zurich, Switzerland.

[6]Centro de Química – Vila Real, Departamento de Química, Universidade de Trás-os-Montes e Alto Douro, 5000–801 Vila Real, Portugal.

[7]Institute of Experimental Physics Slovak Academy of Sciences, Watsonova 47, Košice, Slovak Republic.

[8]European Synchrotron Radiation Facility, 38043 Grenoble, France.

*Corresponding author: rvsilva@fc.up.pt

*Corresponding author: jamoreir@fc.up.pt

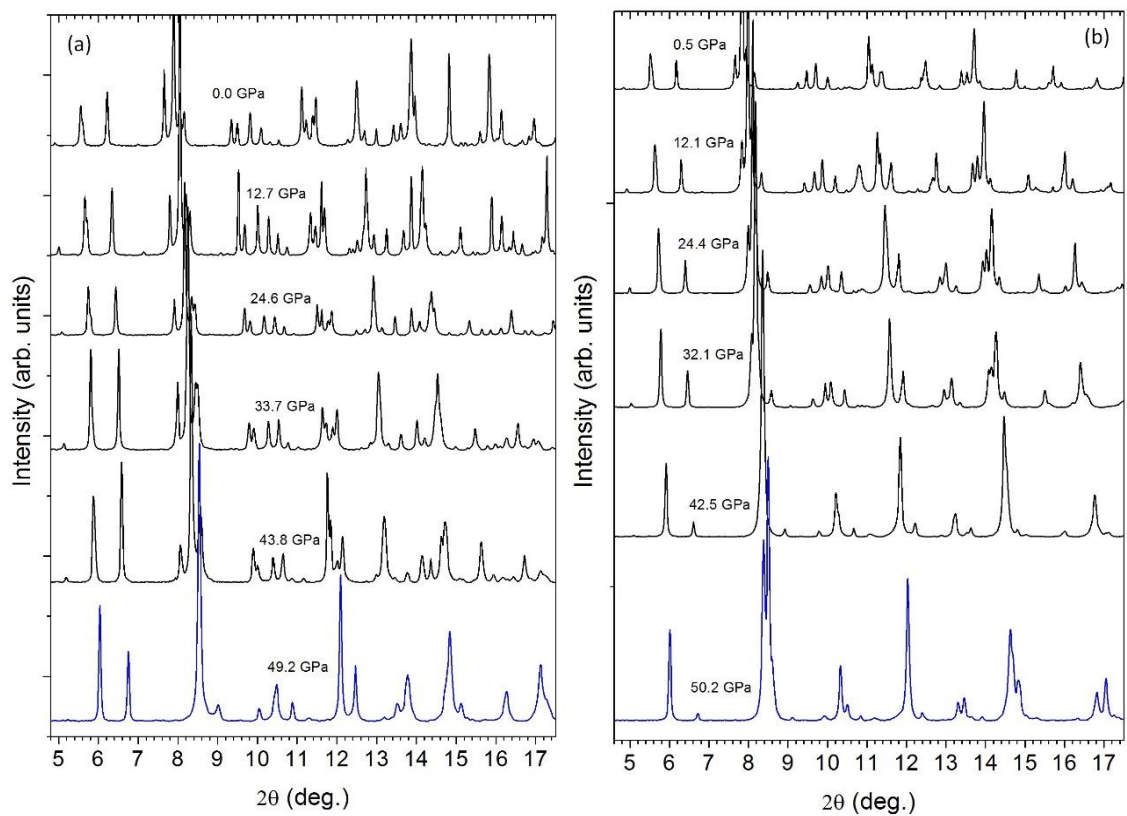

**Figure S1.** Representative XRD patterns of (a) TbFeO$_3$ and (b) SmFeO$_3$ recorded at different fixed pressures.

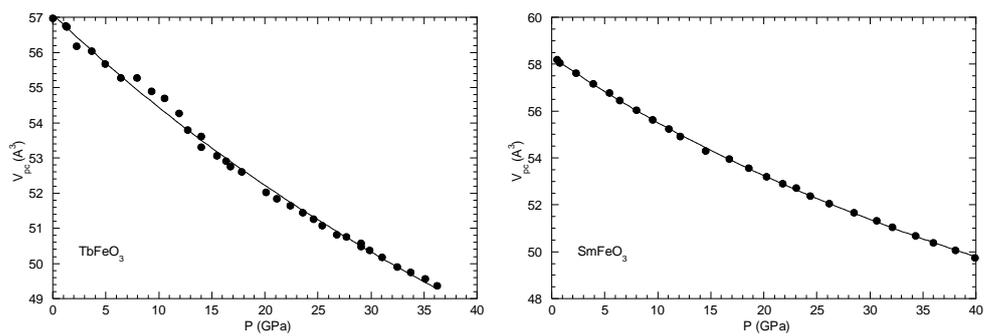

**Figure S2.** Pseudocubic volume of TbFeO$_3$ and SmFeO$_3$ as a function of pressure, in the orthorhombic phase. The solid line was calculated from the best fit of the third-order Birch-Murnaghan equation of state (Eq. 1) to the experimental data.

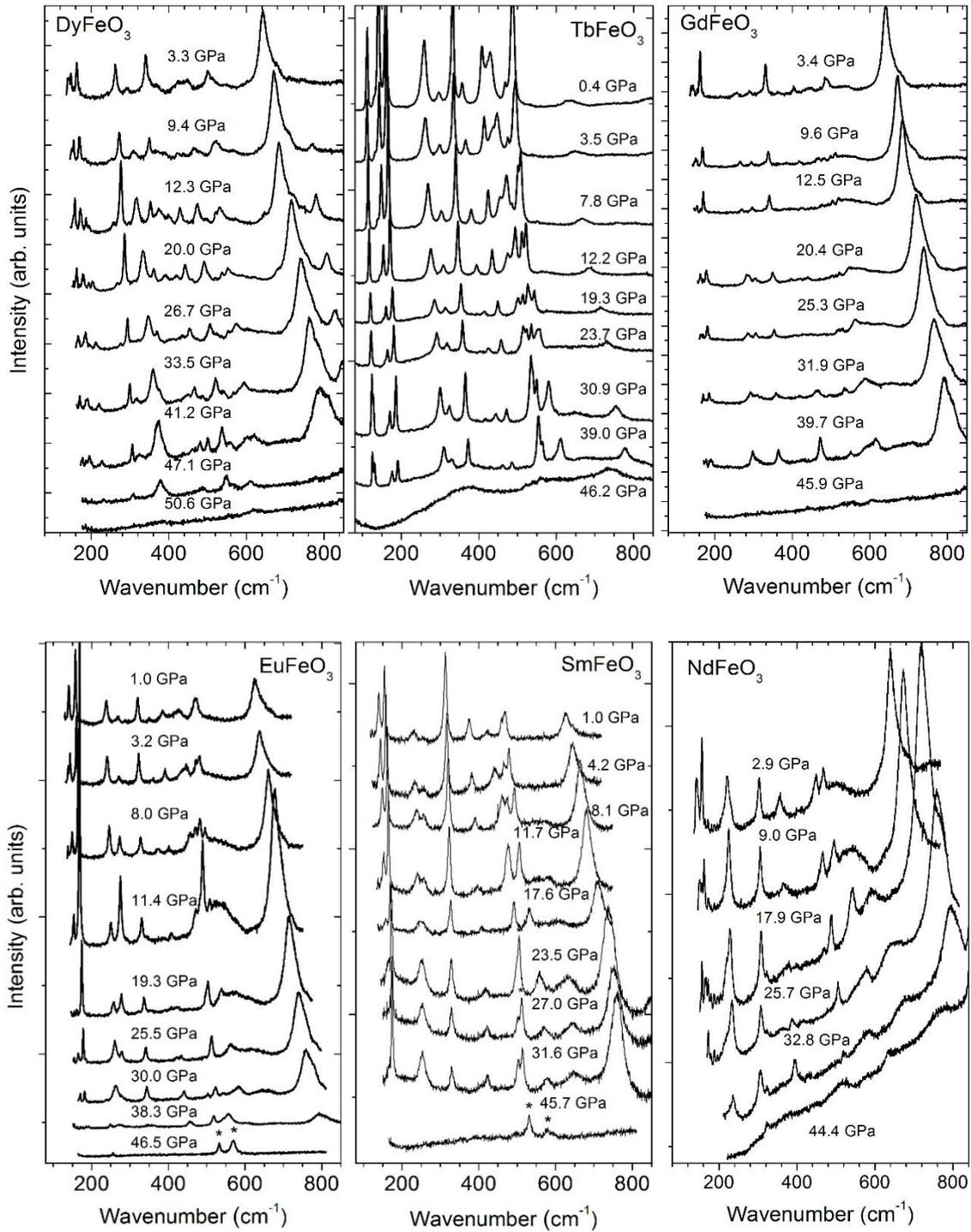

**Figure S3.** Representative Raman spectra of DyFeO$_3$, TbFeO$_3$, GdFeO$_3$, EuFeO$_3$, SmFeO$_3$ and NdFeO$_3$ recorded at different applied pressures. The peaks marked with (*) correspond to Raman modes arising from impurity solid phases, which were unintentionally introduced as a minor phase in the helium transmitting medium in the DAC preparation procedure.

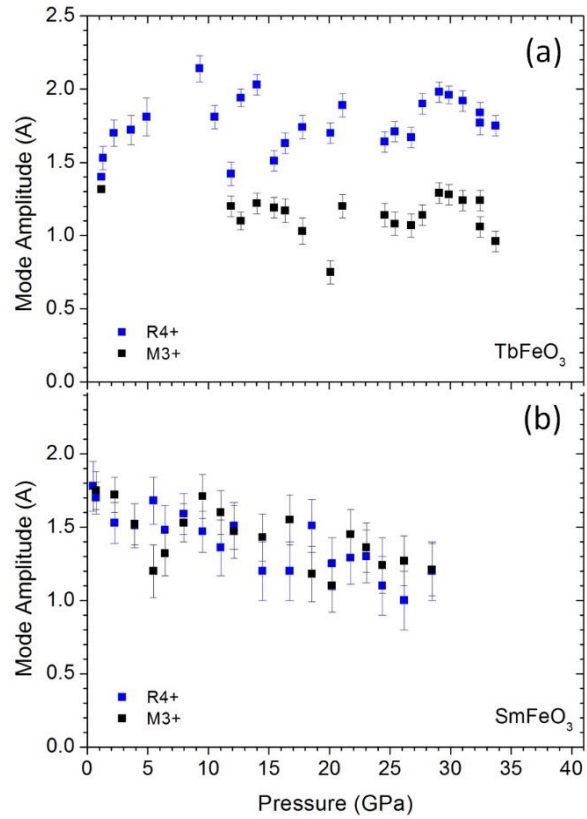

**Figure S4.** R4+ and M3+ distortions amplitude as a function of pressures obtained by Amplimodes refinement of (a) TbFeO$_3$ and (b) SmFeO$_3$ XRD patterns. For the case of SmFeO$_3$, the smallest distortion (M2+, not presented in this figure) was fixed at the constant value of 0.22 Å to reduce scattering.

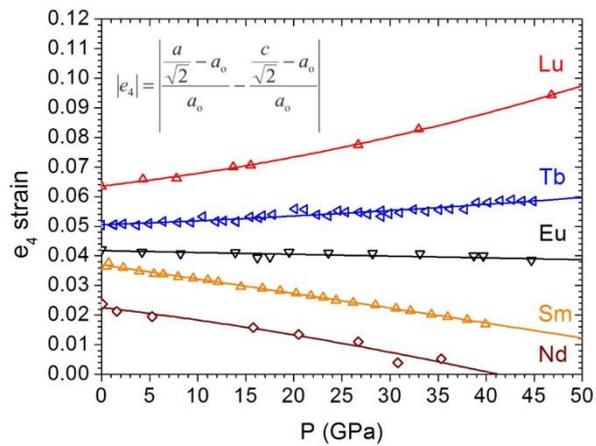

**Figure S5.** Spontaneous $e_4$ strain as a function of pressure for different $R$FeO$_3$. Data of LuFeO$_3$ and EuFeO$_3$ were taken from Ref. 17 and NdFeO$_3$ from Ref. 19.

## I. Pressure dependence of the Raman active rotational modes

For TbFeO$_3$ (SmFeO$_3$), the upper shift (down shift) of the frequency of both rotational modes relatively to the universal linear dependence on the BO$_6$ volume is observed in Figure S6(a) (S6(b)). The observed upper shift (down shift) is a direct consequence of the increase (decrease) of the tilt angles.

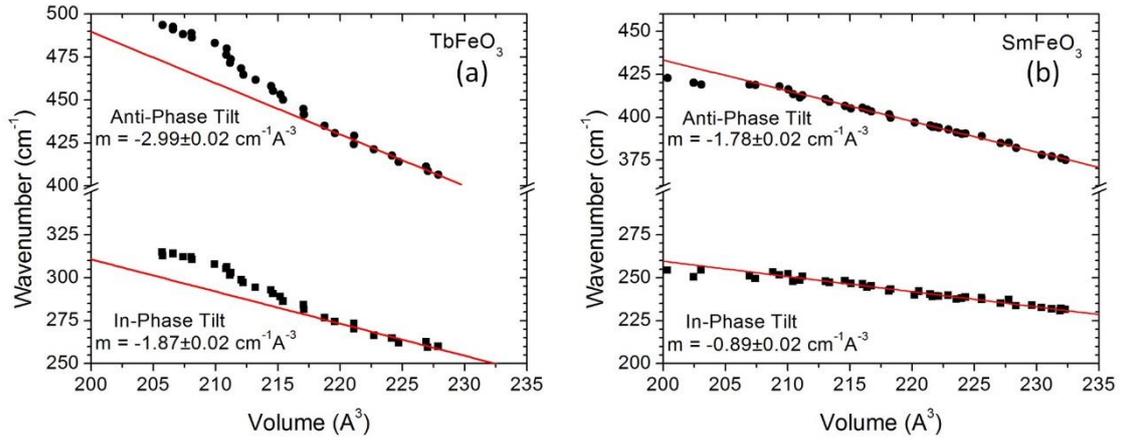

**Figure S6.** Volume dependence of the frequency of the Raman octahedra rotational modes of (a) TbFeO$_3$ and (b) SmFeO$_3$.

**Table I.** Pseudo-cubic lattice parameters as a function of pressure TbFeO$_3$ and SmFeO$_3$.

| TbFeO$_3$ | | | | SmFeO$_3$ | | | |
|---|---|---|---|---|---|---|---|
| P (GPa) | $a_{pc}$ (Å) | $b_{pc}$ (Å) | $c_{pc}$ (Å) | P (GPa) | $a_{pc}$ (Å) | $b_{pc}$ (Å) | $c_{pc}$ (Å) |
| 1.2 | 3.9547(3) | 3.8162(4) | 3.7609(9) | 0.5 | 3.9573(9) | 3.852(9) | 3.8164(7) |
| 1.3 | 3.9541(8) | 3.8154(5) | 3.7600(8) | 0.7 | 3.9569(7) | 3.849(8) | 3.8115(2) |
| 2.2 | 3.9428(1) | 3.8013(1) | 3.7482(8) | 2.3 | 3.9433(9) | 3.840(6) | 3.8042(3) |
| 3.67 | 3.9386(2) | 3.7982(5) | 3.7460(8) | 3.9 | 3.9306(7) | 3.830(7) | 3.7970(2) |
| 4.9 | 3.9307(1) | 3.7914(5) | 3.7362(1) | 5.6 | 3.9197(1) | 3.823(1) | 3.7893(9) |
| 6.4 | 3.9225(5) | 3.7822(6) | 3.7258(8) | 6.4 | 3.9120(7) | 3.8154(5) | 3.7826(7) |
| 7.9 | 3.9216(2) | 3.7829(6) | 3.7263(1) | 7.8 | 3.9005(4) | 3.806(5) | 3.7747(5) |
| 9.3 | 3.9147(6) | 3.7696(6) | 3.7199(6) | 9.5 | 3.8893(7) | 3.797(7) | 3.7657(7) |
| 10.5 | 3.8989(7) | 3.7585(7) | 3.7036(5) | 11.0 | 3.8785(5) | 3.790(1) | 3.7572(8) |
| 11.9 | 3.8892(3) | 3.7489(7) | 3.6936(9) | 12.1 | 3.8699(2) | 3.783(6) | 3.7515(6) |
| 12.7 | 3.8867(9) | 3.7417(8) | 3.6891(4) | 14.5 | 3.8511(2) | 3.770(5) | 3.7395(3) |
| 14.0 | 3.8831(2) | 3.7426(7) | 3.6892(3) | 16.7 | 3.8416(4) | 3.763(2) | 3.7321(1) |
| 15.4 | 3.87420 | 3.7270(1) | 3.6744(6) | 18.6 | 3.8306(1) | 3.7551(5) | 3.7246(1) |
| 16.3 | 3.8668(2) | 3.7252(4) | 3.6649(1) | 20.2 | 3.8199(3) | 3.747(3) | 3.7173(3) |
| 16.7 | 3.8666(3) | 3.72490 | 3.6664(7) | 21.8 | 3.8107(4) | 3.741(1) | 3.7115(3) |
| 17.8 | 3.8650(3) | 3.7230(4) | 3.6641(3) | 23.1 | 3.8047(3) | 3.737(5) | 3.7075(7) |
| 20.1 | 3.8525(5) | 3.7060(3) | 3.6438(8) | 24.4 | 3.7947(6) | 3.7296(5) | 3.7014(9) |
| 21.1 | 3.8490(2) | 3.6994(5) | 3.6413(3) | 26.2 | 3.7844(4) | 3.7232(5) | 3.6940(7) |
| 22.4 | 3.8414(5) | 3.6927(9) | 3.6408(5) | 28.5 | 3.772(7) | 3.7154(5) | 3.6857(9) |
| 23.6 | 3.8341(3) | 3.6906(1) | 3.6344(7) | 30.6 | 3.7622(3) | 3.706(6) | 3.6797(1) |
| 24.6 | 3.8298(2) | 3.6891(4) | 3.6265(7) | 32.1 | 3.7533(9) | 3.701(3) | 3.6741(3) |
| 25.4 | 3.8257(1) | 3.6845(5) | 3.6219(1) | 34.3 | 3.7417(3) | 3.693(5) | 3.6670(6) |
| 26.8 | 3.8191(9) | 3.6799(5) | 3.6163(8) | 36.0 | 3.7325(3) | 3.68(8) | 3.660(9) |
| 27.7 | 3.8171(9) | 3.6763(8) | 3.6142(7) | 38.0 | 3.7212(2) | 3.682(1) | 3.653(2) |
| 29.1 | 3.8114(3) | 3.6716(5) | 3.6071(1) | 39.9 | 3.7106(1) | 3.674(6) | 3.6484(6) |
| 29.9 | 3.8075(5) | 3.6686(3) | 3.6015(5) | 42.5 | 3.602(2) | 3.628(2) | 3.641(6) |
| 31.0 | 3.8027(5) | 3.6642(2) | 3.5958(5) | 44.1 | 3.590(8) | 3.622(8) | 3.636(5) |
| 32.5 | 3.7973(6) | 3.6585(3) | 3.5900(6) | 45.2 | 3.581(8) | 3.613(1) | 3.634(2) |
| 33.7 | 3.7916(1) | 3.6553(7) | 3.5839(2) | 46.4 | 3.562(1) | 3.595(6) | 3.627(6) |
| 35.1 | 3.7872(3) | 3.6499(5) | 3.5785(3) | 47.9 | 3.542(3) | 3.580(6) | 3.619(9) |
| 36.2 | 3.7826(7) | 3.6461(7) | 3.5732(7) | 49.0 | 3.533(8) | 3.568(7) | 3.617(3) |
| 37.7 | 3.7767(1) | 3.6435(8) | 3.5670(6) | 50.2 | 3.522(6) | 3.560(4) | 3.613(3) |
| 39.0 | 3.7724(1) | 3.6408(1) | 3.5603(8) | | | | |
| 40.2 | 3.7681(4) | 3.636(5) | 3.5554(5) | | | | |
| 41.5 | 3.7637(9) | 3.6308(8) | 3.5498(8) | | | | |
| 42.6 | 3.7595(5) | 3.6267(3) | 3.5448(7) | | | | |
| 43.7 | 3.7535(5) | 3.6218(2) | 3.5404(9) | | | | |
| 44.9 | 3.7475(7) | 3.6099(6) | 3.5361(2) | | | | |
| 46.2 | 3.6182(1) | 3.5716(2) | 3.5387(1) | | | | |
| 47.5 | 3.5926(4) | 3.5603(4) | 3.5416(6) | | | | |
| 49.2 | 3.5632(5) | 3.5467(8) | 3.5351(2) | | | | |
| 49.8 | 3.5610(5) | 3.5435(5) | 3.5320(2) | | | | |
| 50.3 | 3.5547(8) | 3.5414(9) | 3.5312(1) | | | | |
| 51.2 | 3.54790 | 3.5382(4) | 3.5276(7) | | | | |
| 52.3 | 3.5418(7) | 3.5360(7) | 3.5245(1) | | | | |
| 53.2 | 3.5373(2) | 3.5342(6) | 3.5211(9) | | | | |